\documentclass[fleqn,10pt]{wlscirep}
\usepackage[utf8]{inputenc}
\usepackage[T1]{fontenc}
\usepackage{multirow}
\usepackage{multicol}
\usepackage{booktabs}
\usepackage{subfig}
\usepackage{graphicx}
\usepackage{amsmath,amsfonts,amsopn,amssymb}
\usepackage{lmodern}
\usepackage{gensymb}
\usepackage{times}
\usepackage{color,soul}
\usepackage{float}
\usepackage{placeins}
\usepackage{siunitx}
\usepackage{ulem}
\usepackage[parfill]{parskip}
\DeclareUnicodeCharacter{2009}{\,} 

\usepackage{mathastext}
\usepackage{wrapfig}
\usepackage{xcolor}
\usepackage{hyperref}
\usepackage{pdfpages}

\newcommand{\sulfacid}{H\textsubscript{2}SO\textsubscript{4}}

\newcommand\tsc[1]{\textsubscript{#1}}
\newcommand\tSc[1]{\textsuperscript{#1}}

\usepackage{setspace}
\usepackage{bm}
\title{Imaging the Breathing of a Platinum Nanoparticle in Electrochemical Environment}

\author[1,2,3]{Clément Atlan}
\author[1,2]{Corentin Chatelier}
\author[2]{Isaac Martens}
\author[1,2]{Maxime Dupraz}
\author[3]{Arnaud Viola}
\author[1,2]{Ni Li}
\author[4]{Lu Gao} 
\author[2]{Steven J. Leake}
\author[2]{Tobias U. Schülli}
\author[1]{Joël Eymery} 
\author[3,*]{Frédéric Maillard} 
\author[1,2,*]{Marie-Ingrid Richard}

\affil[1]{Univ. Grenoble Alpes, CEA Grenoble, IRIG, MEM, NRS, 17 rue des Martyrs, F-38000 Grenoble, France}
\affil[2]{ESRF - The European Synchrotron, 71 Avenue des Martyrs, F-38000 Grenoble, France}
\affil[3]{Univ. Grenoble Alpes, Univ. Savoie Mont Blanc, CNRS, Grenoble INP, LEPMI, 38000 Grenoble, France}
\affil[4]{Laboratory for Inorganic Materials and Catalysis, Department of Chemical Engineering and Chemistry, Eindhoven University of Technology, P. O. Box 513, 5600 MB Eindhoven, The Netherlands}

\affil[*]{E-mails: (F.M.) frederic.maillard@grenoble-inp.fr, (M.-I.R.) mrichard@esrf.fr}

\begin{abstract}
Surface strain is widely employed in gas phase catalysis and electrocatalysis to control the binding energies of adsorbates on active sites. However, \textit{in situ} or \textit{operando} strain measurements are experimentally challenging, especially on nanomaterials. Here, we exploit coherent diffraction at the new 4\tSc{th} generation Extremely Brilliant Source at the European Synchrotron Radiation Facility (ESRF-EBS) to map and quantify strain inside individual Pt catalyst nanoparticles under electrochemical control. 3D nanoresolution strain microscopy together with density functional theory and atomistic simulations show for the first time evidence of heterogeneous and potential-dependent strain distribution between highly-coordinated (\{100\} and \{111\} facets) and under-coordinated atoms (edges and corners) as well as evidence of strain propagation from the surface to the bulk of the nanoparticle. These dynamic structural relationships directly inform the design of strain-engineered nanocatalysts for energy storage and conversion applications.
\end{abstract}
\begin{document}

\flushbottom
\maketitle

\thispagestyle{empty}


Controlling the degree of surface strain is commonly used to optimise the chemisorption energies of adsorbates onto metals, and thus accelerate the rate of (electro)catalytic reactions. Strain is the distortion of a crystal lattice due to a localised or a global stress, which displaces the atoms from their equilibrium positions and alters the surface's electronic \textit{d}-band centre\cite{hammer_electronic_1995}. Surface strain can be created by (i) depositing an overlayer of one metal onto another metal with a different lattice parameter\cite{yang_platinum-based_2011}, (ii) by alloying the metals\cite{stamenkovic_trends_2007,stamenkovic_improved_2007,strasser_lattice-strain_2010, greeley_alloys_2009}, (iii) introducing structural defects\cite{bandarenka_elucidating_2014, dubau_defects_2016,chattot_beyond_2017,chattot_surface_2018} or (iv) simply controlling the size and/or the shape of nanomaterials\cite{sneed_building_2015}. The fundamental relationship between surface strain and the rate of a (electro-)catalytic reaction was pioneered by Hammer and Nørskov using density functional theory (DFT) calculations\cite{hammer_electronic_1995,hammer_why_1995}. They showed that compressive strain facilitates orbital overlap, broadening the band structure and downshifting the \textit{d}-band centre, thereby weakening the binding of adsorbates. The \textit{d}-band model has become ubiquitous in the fields of energy storage and conversion systems (\textit{e.g.}, proton-exchange membrane fuel cells and water electrolyzers) and in the production of high value added products (H\tsc{2}O\tsc{2} and Cl\tsc{2}). This theory predicted that the rate of the oxygen reduction reaction (ORR) would be enhanced on catalysts which bind hydroxyl species (*OH) \textit{ca.} 0.10 to 0.15 eV weaker than Pt(111)\cite{hammer_electronic_1995,hammer_why_1995}, later experimentally verified by Stamenkovic \textit{et al.} using a Pt\tsc{3}Ni(111)-skin surface\cite{stamenkovic_improved_2007}. Nevertheless, transferring the performance obtained on perfect single crystals to nanocatalysts proved challenging, in part because nanocatalyts present multiple catalytic sites with a range of binding energies. By introducing the “generalized coordination number” ($\overline{CN}$)\cite{calle-vallejo_finding_2015}, Calle-Vallejo \textit{et al.} successfully extended the structure binding-energy relationship for ORR to stepped Pt surfaces\cite{bandarenka_elucidating_2014}, defective Pt(111) surfaces\cite{calle-vallejo_why_2017} and to Pt nanoparticles (NPs)\cite{calle-vallejo_why_2017,ruck_oxygen_2018}. They predicted that locally concave sites ($7.5 < \overline{CN}  \leq 8.3$) are more active towards the ORR than terrace Pt(111) sites ($\overline{CN} = 7.5$) and that locally convex or buried sites ($ \overline{CN} < 7.5$ or $\overline{CN} > 8.3$) contribute poorly to the global ORR rate because they strongly or weakly bind the ORR intermediates. An important corollary of this conceptual framework is that defective surfaces, featuring various $\overline{CN}$ numbers, may break the scaling relationships between reaction intermediates, and thus accelerate the rate of multiple electrocatalytic reactions\cite{abild-pedersen_scaling_2007,busch_beyond_2016}. This behavior has been verified for the ORR, CO oxidation, and methanol/ethanol oxidation reactions on PtNi/C nanocatalysts\cite{chattot_beyond_2017}.\cite{chattot_surface_2018} Several spectroscopic and diffraction methods have emerged for measuring strain distributions averaged over billions of NPs.\cite{chattot_surface_2018} However, surface \textit{vs} bulk localised strain, interactions between active sites, and the role of competitively adsorbing species are still experimentally absent and challenging to model. An \textit{in situ} understanding of how strain is distributed over a single NP and its dependence on electrode potential is therefore needed to connect these average strain distributions to specific structural motifs.



Here, we investigate these issues on an individual Pt NP using recent advances in Bragg Coherent Diffraction Imaging (BCDI)\cite{robinson_coherent_2009,ulvestad_situ_2016,passos_three-dimensional_2020,hua_structural_2019,choi_situ_2020,ulvestad_single_2014,bjorling_coherent_2019} and the 4\tSc{th} generation Extremely Brilliant Source of the European Synchrotron Research Facility in Grenoble, France. Our results reveal that strain is heterogeneously distributed between highly- and weakly-coordinated surface atoms, and propagates from the surface to the bulk of the Pt NP as the electrode potential increases in the double layer region ($0.26 \leq \textit{E} \leq 0.56$ V).

\begin{figure*}[h!]
    \centering
    \includegraphics[width=1\linewidth]{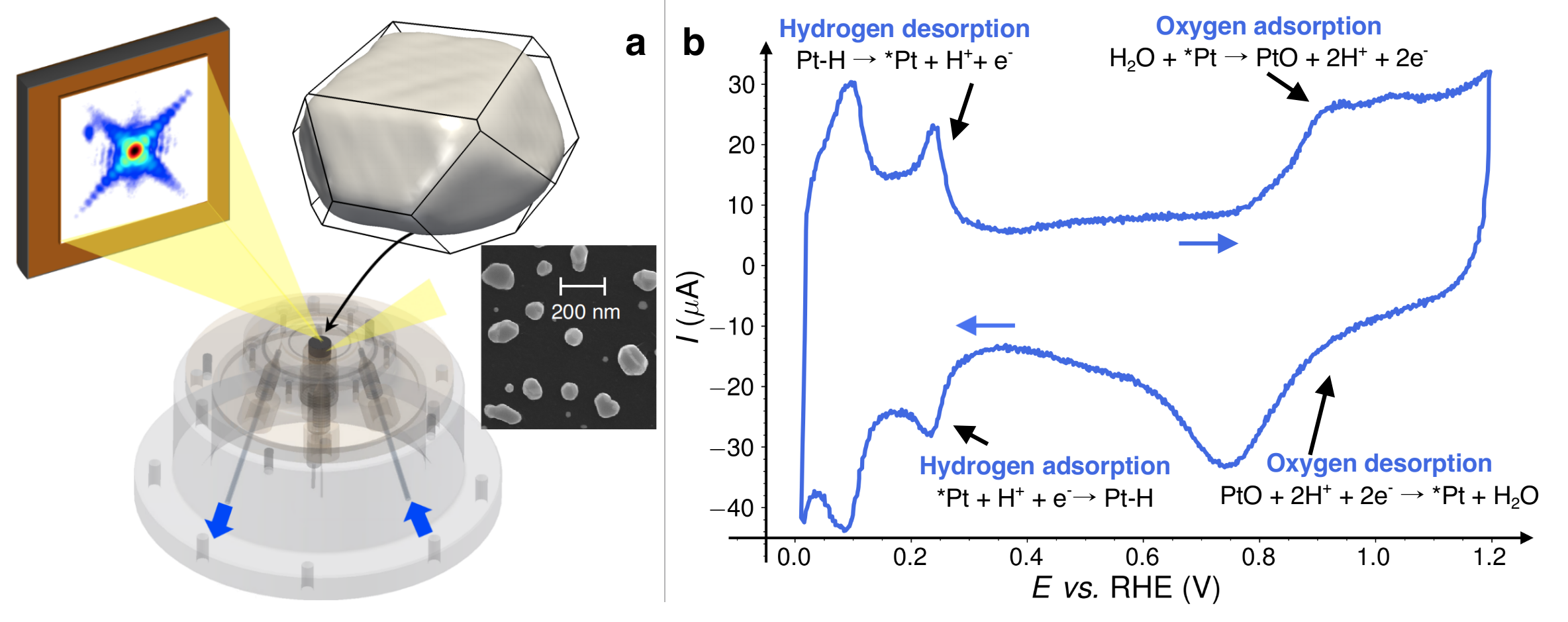}
    \caption{\textbf{Scheme of the experimental setup to image strain distribution onto a single Pt nanoparticle.} \textbf{a}, Drawing of the \textit{in situ} BCDI flow cell, of the X-ray beam and of the two-dimensional detector and view of the reconstructed Pt nanoparticle. \textbf{b}, Cyclic voltamogram of the Pt/GC electrode measured in the \textit{in situ} BCDI flow cell, \textit{T} = 24 $\pm$ 0.1$^{\circ}C$, \textit{v} = 50 mV s\tSc{-1},  0.05 M H$_2$SO$_4$.}
    \label{fig:exp_scheme}
\end{figure*}

\begin{figure*}[h!]
    \centering
    \includegraphics[width=0.65\linewidth]{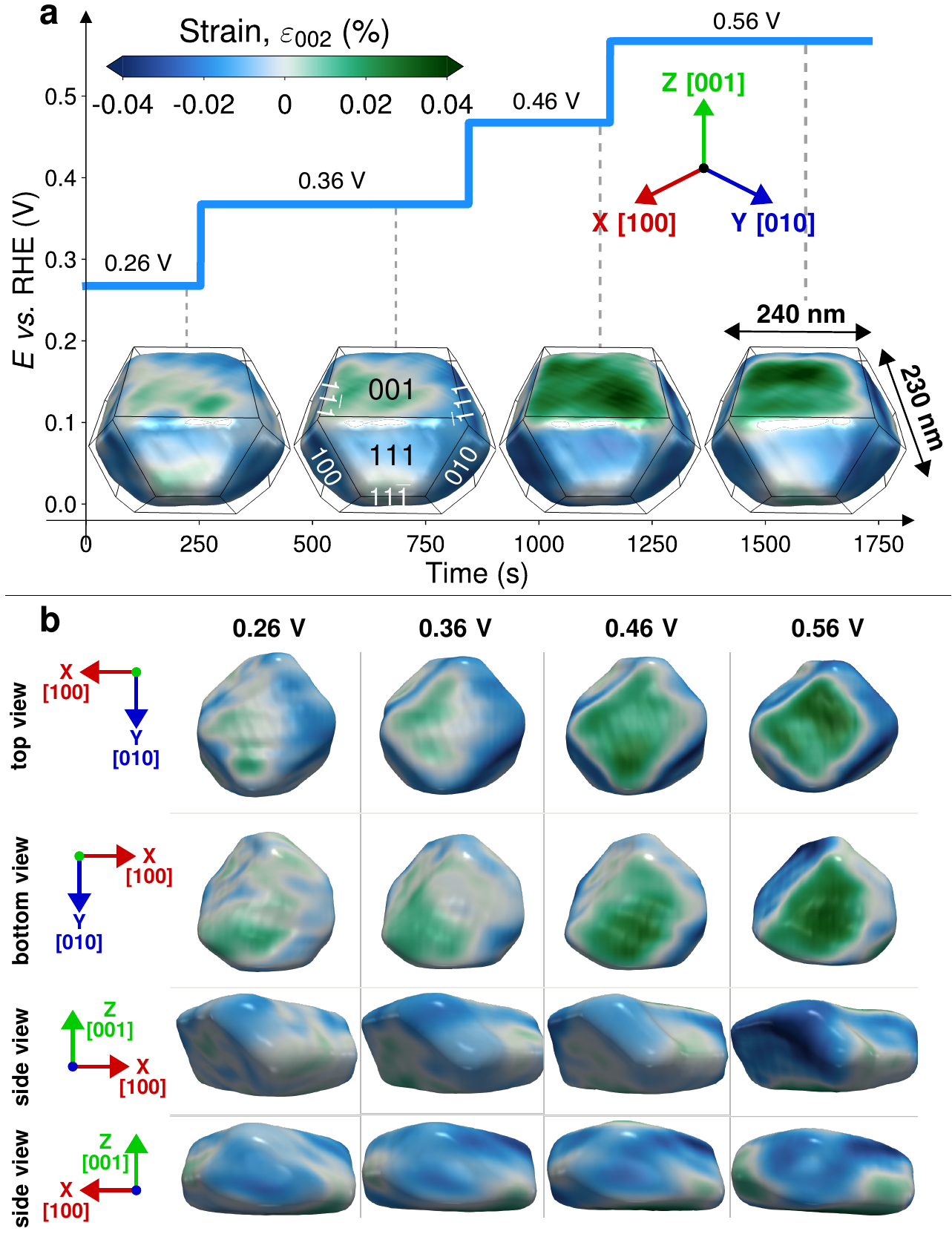}
    \caption{\textbf{Revealing the influence of the electrode potential on the surface strain ($\varepsilon_{002}$).} Panels \textbf{a} and \textbf{b} present the same particle viewed from different perspectives. The colour bar is identical across the panels \textbf{a} and \textbf{b}. See Supplementary Figure 4 for multiple views of the reconstructed phase at different electrode potentials.} 
    \label{fig:step_recons}
\end{figure*}

The Pt NPs supported on glassy carbon (Pt/GC) used in this study were obtained by dewetting a 30 nm Pt film at 750$^{\circ}C$ for 10 hours under Ar atmosphere. As shown in Supplementary Figure 1, their size ranges from 50 nm to 500 nm, providing a useful platform to investigate changes in morphology and surface strain over NPs of different sizes and crystallographic orientations. Figure \ref{fig:exp_scheme} displays a schematic of the experimental setup and a cyclic voltammogram (CV) of the Pt/GC NPs in 0.05 M H\tsc{2}SO\tsc{4} electrolyte. The CV displays the conventional features associated with the adsorption and desorption of under-potentially-deposited hydrogen (H\tsc{upd}, $0.05 \leq \textit{E} \leq 0.30$ V), and the formation and reduction of Pt surface oxides ($\textit{E} \geq 0.80$ V).
Using a focused beam of coherent X-rays, BCDI was employed to retrieve the morphology and the projected three-dimensional displacement field within a single Pt NP at different electrode potentials. Since NPs experience strain from surface tension, their atoms are locally distorted from their ideal positions. The reconstructed electron density is a complex number, with an amplitude reflecting the particle morphology and the phase corresponding to the projection of the displacement field onto the scattering vector. This information is encoded into reciprocal space in the form of a fringed diffraction pattern (see Supplementary Figure 2 and Methods). Phase retrieval algorithms (see Supplementary Figure 3 and Methods) are then used to invert the data from reciprocal space to direct space, and to construct maps of displacement and strain as a function of the electrode potential.
A reconstructed Pt NP, \textit{ca.} $240\times230\times110$ (height) nm\tSc{3} in size, is displayed in Figure \ref{fig:step_recons} and in the inset of Figure \ref{fig:exp_scheme} (voxel size of $5 \times 5 \times 5$ nm\tSc{3}, the spatial resolution is estimated at 18 nm, see Supplementary Figure 5). The retrieved electron density map is drawn at an isosurface of 65 \% of its maximum, as suggested by Ref.\cite{carnis_towards_2019}. The \textit{z}-axis is drawn vertically and corresponds to the [001] crystallographic direction of the NP. The \textit{x}-axis and \textit{y}-axis correspond to the [100] and [010] crystallographic directions of the NP, respectively. The same crystallographic (\textit{x}, \textit{y}, \textit{z}) frame is used throughout this study. By calculating the angles between the different facets, we identified 8 facets with \{111\} orientation, and 5 facets with \{100\} orientation, the top facet being a (001) facet. The geometrically necessary 6\tSc{th} \{100\} facet is very small and was not recognised automatically by our algorithm (see Methods for the facet analysis). All facets are indexed by their Miller indices (hkl) in Figure \ref{fig:step_recons} and in Supplementary Figure 6.
Since the \textbf{002} Bragg reflection was measured, we refer to "strain" as the heterogeneous strain along the [001] direction (along the NP’s height, the scattering vector \textbf{\textit{Q}}\tsc{002} being parallel to [001]\cite{robinson_coherent_2009}, see Methods) computed as the derivative of the [001]-projected displacement $\textit{u}_{002}$ varying along the same axis, as follows:
\begin{equation}
    \varepsilon_{002} = \frac{\partial{\textit{u}_{002}}}{\partial{\textit{z}_{[001]}}} = \frac{\partial{\textit{u}_{002}}}{\partial{\textit{z}}}.
    \label{eq:strain}
\end{equation}
Notice that the above-defined strain in equation \ref{eq:strain} (called merely "strain") is rigorously the \textit{heterogeneous} strain that differs from the \textit{homogeneous} strain defined further. As we strive to quantify strain in different regions of the Pt NP, the voxels belonging to the bulk, the surface, the facet or the edge and corner atoms have been segmented. The definitions of the terms "surface strain", "bulk strain", "strain at edge and corner atoms", "facet strain" and "global strain", can be found in the Methods section.

Figure \ref{fig:step_recons}a displays changes in surface strain over the potential window $0.26 < \textit{E} < 0.56$ V. 
In agreement with previous work\cite{Hartl2010,Dubau2013}, we repeatedly observed that Pt NPs become mobile on the substrate at \textit{E} > 0.6 V, prohibiting BCDI measurements at higher potentials. The precise mechanism remains unclear but is likely caused by Pt-catalysed corrosion of the GC support. Different views are shown in Figure \ref{fig:step_recons}b. Regardless of perspective, positive strain (tension) accumulates on the top and bottom facets whereas edges, corners, and side facets become negatively strained (compression) as the electrode potential increases. These trends are quantified in Supplementary Figure 7, where the averaged strain is plotted for specific regions of the NP. Side facets, edges, and corners exhibit a compressive strain with increasing potential, whereas the top (001) and bottom ($00\overline{1}$) facets accumulate tensile strain. Electrocompressibility with applied potential has been previously measured on planar surface with \textit{in situ} surface diffraction\cite{melroy_two-dimensional_1988}. But electrocompressibility can not account for such observed heterogeneity in strain distribution. 

\begin{figure*}
  \centering
  \includegraphics[width=1\linewidth]{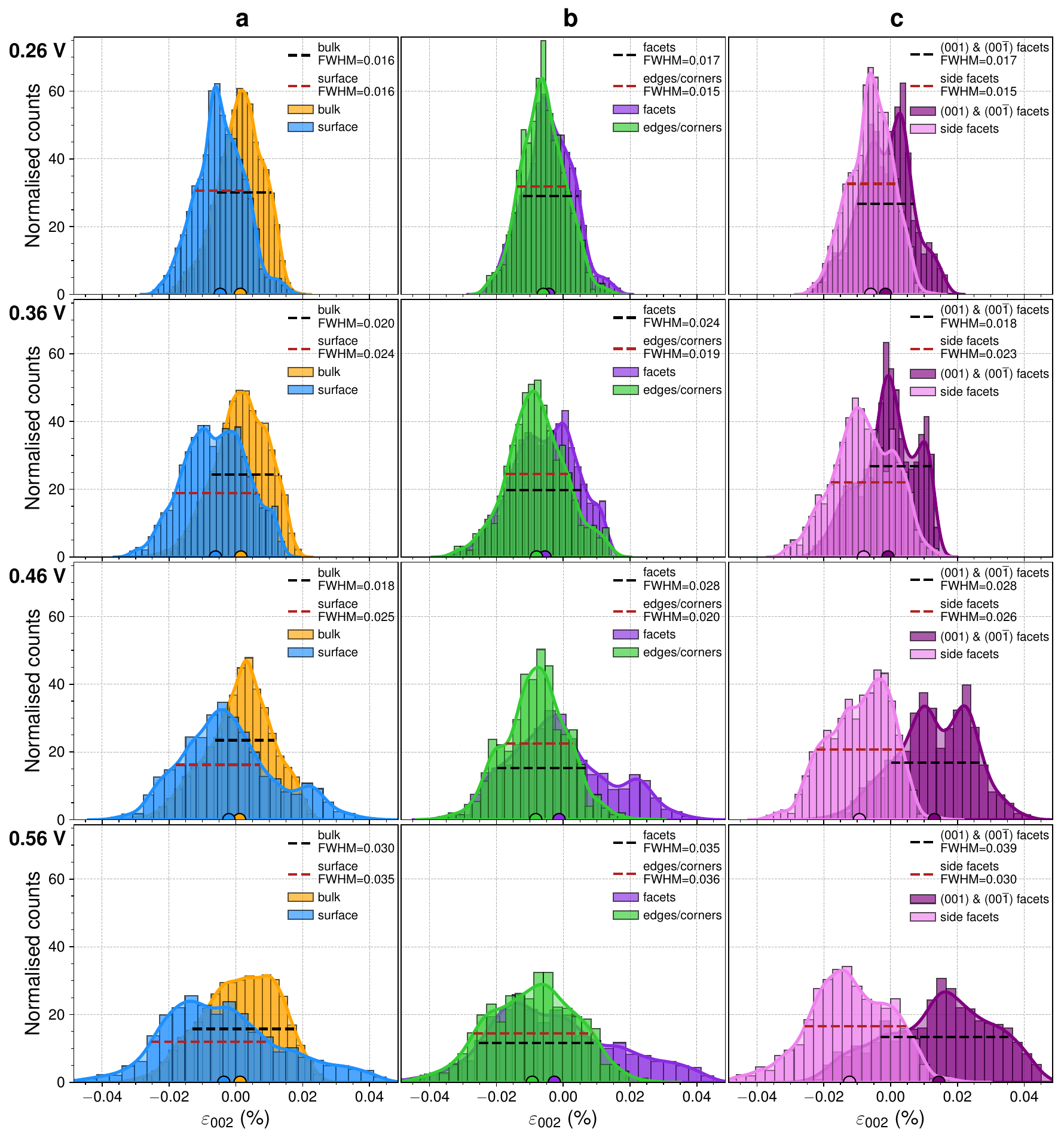}
  \caption{\textbf{Distributions of surface \textit{vs.} bulk strain, facet \textit{vs.} edge/corner strain and top/bottom facet \textit{vs.} side facet strain as a function of the electrode potential.} The bulk and surface strain distributions are represented in blue and yellow in panel \textbf{a} column, respectively. Note that the number of voxels belonging to the bulk is much larger than the number of voxels belonging to the other regions; however, all histograms have been normalised. This methodology allowed calculating the probability density function (continuous lines). The number of bins was held constant across different electrode potentials by altering the bin size. As the potential increases, the full width at half maximum of the surface strain increases more than that of the bulk strain, indicating that heterogeneity in strain distribution initiates at the surface, and then propagates into the bulk of the Pt nanoparticle.
  The facet and edge/corner strain distributions are represented in violet and green in panel \textbf{b}, respectively. Panel \textbf{c} displays the distributions of the side facets and that of the top and bottom facets in pink and purple respectively, highlighting the bimodal distribution of the facet strain in panel \textbf{b} in violet. See Supplementary Figure 8 for the global heterogeneous strain distribution.}
  \label{fig:histo_all_regions}
\end{figure*}

The strain histograms shown in Figure \ref{fig:histo_all_regions} reveal additional layers of heterogeneity. At 0.26 V, the histograms of the strain distribution possess a sharp, Gaussian shape, almost zero-centred (Fig. \ref{fig:histo_all_regions}, top row, also Supplementary Figure 7). In Figure \ref{fig:histo_all_regions}a (blue color), the full width at half maximum (FWHM) of the surface strain distribution broadens from 0.016 \% at 0.26 V to 0.035 \% at 0.56 V. This is because the different surface regions have opposite strain contributions, as mentioned above. In contrast, the FWHM of the bulk strain distribution (yellow color) remains close to 0.020 \% between 0.26 V and 0.46 V, but experiences a sharp increase to 0.030 \% at 0.56 V, which suggests that the bulk region of the Pt NP accommodates the strain which develops on the surface. The edges and corners (Fig. \ref{fig:histo_all_regions}b, green color) shift left, indicating compression, while simultaneously broadening as the potential increases. The facet voxels (Fig. \ref{fig:histo_all_regions}b, purple color) do not show substantial directional shift but exhibit extreme broadening, and even a bimodal distribution at 0.56 V.
This bimodal facet strain is decomposed in Figure \ref{fig:histo_all_regions}c. Side facets (in light pink) compress while top and bottom facets (in dark purple) expand. The cooperative compression of edge/corner atoms and of the side facets, and tension of the top and bottom facets (see Figure \ref{fig:step_recons}c) suggest that both effects originate from the same phenomenon. Since only (bi)sulfate ions were present in the electrolyte and since the potential of zero total charge (PZTC) of polycrystalline Pt NPs is close to 0.26 V in sulfuric acid\cite{chen_potential_2010,chattot_electrochemical_2021}, the increase in surface strain heterogeneity measured between 0.26 and 0.56 V is attributed to adsorption of (bi)sulfate ions on the surface, as supported by former studies using scanning tunnelling microscopy, infrared spectroscopy, \textit{etc})\cite{ito_structures_2008,faguy_situ_1996,shingaya_model_1998}. Complementary to these techniques, \textit{in situ} BCDI provides a direct and unique view of how strain is distributed at the NP’s surface and how it responds as a function of the electrode potential.

Supplementary Figure 9a shows the variations of the homogeneous strain which is given by the relative variations of the average \textit{d}-spacing, $\textit{d}_{002}$, computed from the position of the maximal intensity of the Bragg peak. The \textit{d}-spacing at 0.26 V (\textit{d}$_{002,~0.26V}$) provides the reference for the homogeneous strain calculation which is given by the following relation:
\begin{equation}
  \textit{e}_{002} = \frac{\textit{d}_{002} - \textit{d}_{002,~0.26V}}{\textit{d}_{002,~0.26V}}.
\end{equation}

Supplementary Figure 9 provides an overview of the strain evolution by combining the homogeneous strain \textit{e}$_{002}$ and the minimum, maximum and standard deviation values of the heterogeneous strain $\varepsilon_{002}$. Panels a and b show a global increase of the \textit{d}-spacing as a function of the electrode potential which is in line with the recent findings of Martens\cite{martens_probing_2019} and Chattot \textit{et al.} \cite{chattot_electrochemical_2021} showing that the lattice parameter of Pt NPs (\textit{ca.} 2 nm in size) increases upon adsorption and is minimised close to their PZTC. The presented results are distinct from those found on Pt single crystals using surface X-ray diffraction. Markovic and Ross reported + 1.5 and + 2.5 \% increase of the \textit{d}-spacing in the H\tsc{upd} region for Pt(111) and Pt(100) single crystal surfaces in 0.05 M \sulfacid, respectively\cite{markovic_surface_2002}. However, no change was noticed in the so-called “double layer region” ($0.4 < \textit{E} < 0.6$ V) even though (bi)sulfate anions are known to strongly adsorb on these surfaces in this potential range. In contrast, the \textit{in situ} BCDI reconstructions (Figure \ref{fig:step_recons}) and the histograms of the strain distribution (Figure \ref{fig:histo_all_regions}) reveal that strain (negative or positive) builds up at the surface and then propagates into the bulk of the Pt NP. A simple explanation for these discrepancies is the presence of under-coordinated atoms at the nanocatalyst surface.
Indeed, molecular dynamics simulations (MDS) by Wu \textit{et al.} showed that under-coordinated atoms are intrinsically stabilised \textit{via} inward displacement (compression), while atoms located at the facets likely experience outward displacements (tensile strain) for both icosahedral and cuboctahedral Pt nanostructures\cite{wu_icosahedral_2012}. Other BCDI studies on noble metal nanocrystals have shown that compression is mainly localised at vertices and edges, with weaker strain on flat surfaces, while the bulk remains relatively strain free upon adsorption.\cite{kim_curvature_induced_2014,watari_differential_2011,kim_active_2018}

\begin{figure*}[h!]
  \centering
  \includegraphics[width=1\linewidth]{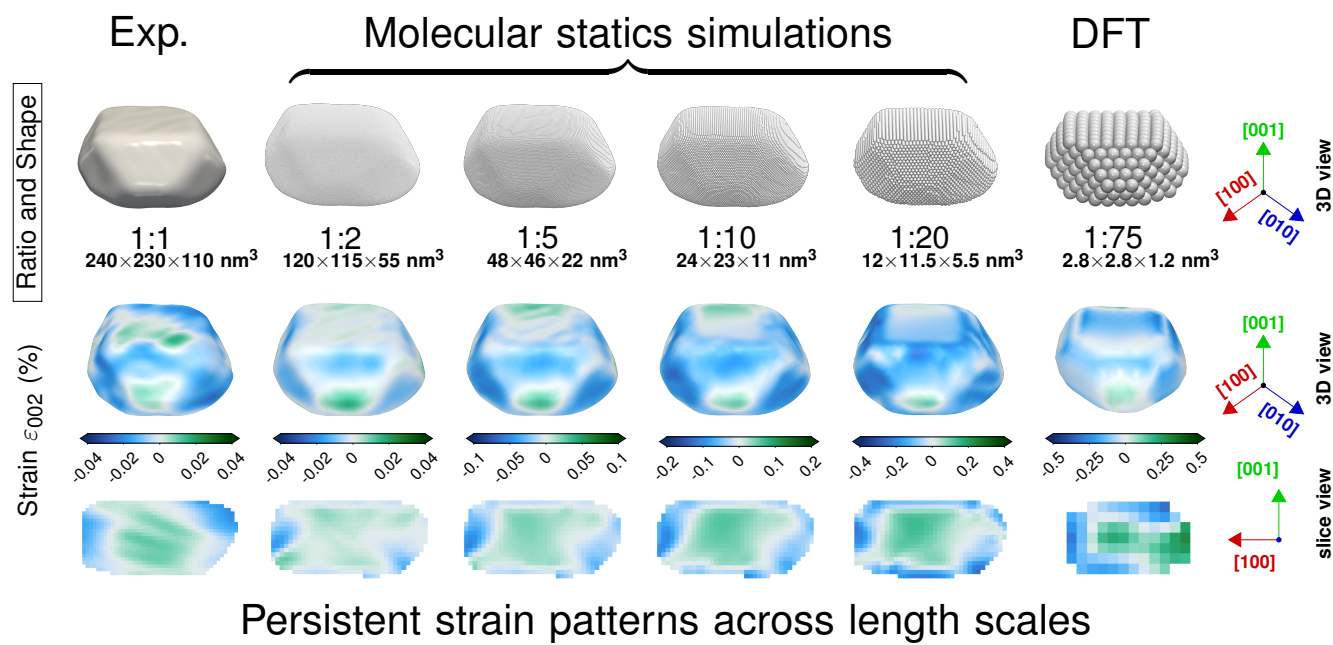}
  \caption{\textbf{Modelling indicates strain gradients are largely independent of crystallite size.} The experimental data (left column) was compared with five model particles of progressively smaller size. For each system, a 3D shape, a 3D strain map and a 2D strain cross section are displayed on each row. The strain colour bar is different for each system because of the different sizes. For the MS simulations, the number of voxels is of the same order of magnitude as that of the experiment. In the DFT simulation, the resolution is nearly atomic. BCDI-equivalent phase maps for each model are available in the Supplementary Figure 10.}
  \label{fig:simulations}
\end{figure*}

Unfortunately, BCDI of practical Pt/C electrocatalysts with 2-5 nm diameter requires flux densities well beyond the reach of current beamlines. Therefore, it is critically necessary to validate to what extent the strain distributions observed in our model system (240 x 230 x 110 nm) apply towards smaller, industrially relevant electrocatalysts. Five nanocrystals with progressively smaller size but identical morphology were modelled using molecular static simulations (MSS) and density functional theory (DFT) calculations. Figure \ref{fig:simulations} shows the strain distributions of six nanocrystals: the experimentally probed particle (at 0.26 V), four particles of decreasing size simulated and relaxed with MSS (ratios 1:2-1:20), and one ~2.5 nm particle  simulated with DFT calculations (ratio of 1:75). The $\varepsilon_{002}$ strain pattern observed by BCDI is largely reproduced in the simulated nanocrystals, without any input beyond morphology. The particle shape produces a compressive strain near the corners and a tensile strain in the bulk of all particles. As expected, the magnitude of the strain is different across the six systems, since the different sizes possess different surface area to volume. These calculations highlight the strong dependence of the strain on the morphology of the nanocrystal, mostly independent of size. Furthermore, the reasonable agreement between MSS and DFT-relaxed nanocrystals demonstrates that the strain effects can be accessed with computationally convenient levels of theory.

\begin{figure*}[h!]
  \centering
  \includegraphics[width=1\linewidth]{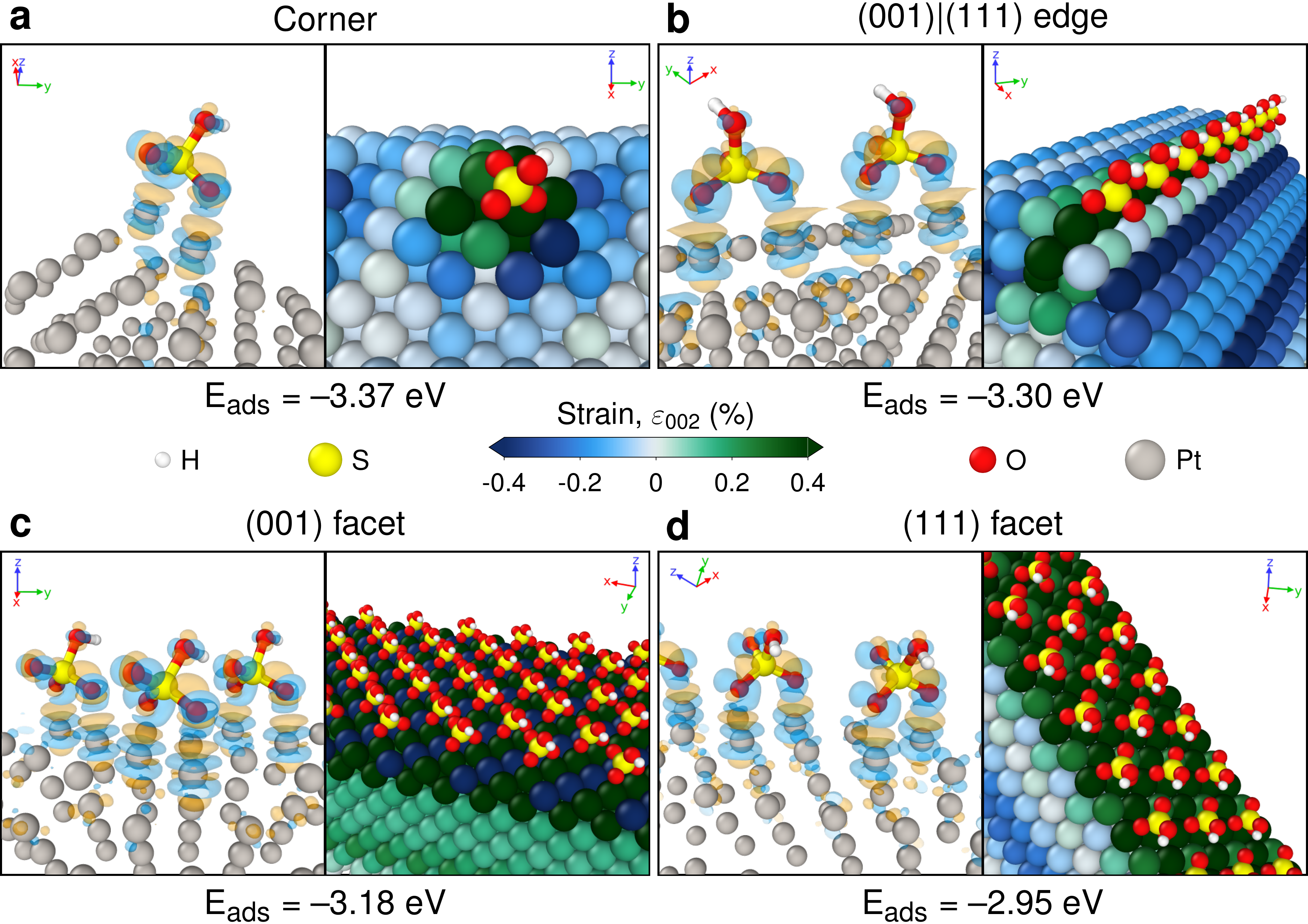}
  \caption{\textbf{Adsorption configurations of HSO\textsubscript{4} on Pt model morphological features (periodic slab modes) and their associated strain response along the [002] direction.} For each panel the left inset displays the isosurface charge density difference ($\pm 0.02~e^{-}/ \si{\angstrom}^{3}$). Negative (respectively positive) isosurfaces are plotted in blue (respectively orange) and correspond to electron-donating regions (respectively electron-accepting regions). The right insets show the strain response $\varepsilon_{002}$ resulting from the associated adsorption. The reference state for each strain calculation is its prior relaxed feature. Panels \textbf{a}, \textbf{b}, \textbf{c} and \textbf{d} correspond to corner, (001)|(111) edge, (001) and (111) facet regions respectively, and are ordered by adsorption energy (more adsorbing to less adsorbing states). The frame axes are the same as the experimental figures axes. See Methods and Supplementary Figures 11 and 12 for more details.}
  \label{fig:adsorption}
\end{figure*}

Additional DFT calculations were performed on a slab model to predict which sites preferentially adsorb bisulfate ions, and understand the atomistic origin of the strain dynamics. Note that only bisulfate adsorption is considered according to the work of Santana \textit{et al.} on Pt(100) surface\cite{santana_dft_2020}. As shown in Figure \ref{fig:adsorption}, corner and edge atoms feature the most favourable adsorption energies (-3.37 and -3.30 eV, respectively) followed by that of the facets: $E_{ads}^{corner} < E_{ads}^{edge} < E_{ads}^{(001)} < E_{ads}^{(111)}$. Whatever the isosurface, the charge density difference predicts that HSO\textsubscript{4} is adsorbed bridging two Pt atoms \textit{via} two Pt-O bonds. In the vicinity of the adsorption sites, atoms experience tension ($\varepsilon_{002} > 0$), except for the free (001) facet atoms. For the corner and edge features, the neighbouring atoms accommodate strain and undergo compression ($\varepsilon_{002} < 0$). All subsurface atoms of the (001) slab experience tension.  Finally, (111) facet surface atoms show tensile behaviour and subsurface atoms mostly undergo compression. The propagation of the compressive strain accommodation is inhibited by the two last atomic layers of the slab, which were fixed to the bulk lattice parameter. Deeper slabs would have resulted in the further propagation of the compression for corner and edge features and tension for the (001) facet. While atomic scale surface details are not visible in BCDI, the technique nevertheless accurately accounts for the changes in strain at a larger scale, where one voxel averages displacement over a few slabs. The observed experimental trend is supported with DFT calculations, showing compression at corners and edges and tension at (001) facets tension. The complex strain behaviour of  the \{111\} facets is also reproduced by DFT, and can be rationalised by considering their relative orientations with respect to the [002] direction. Each slab has been oriented so that the simulated corner, edge, (001) and (111) facets correspond to a top corner, a top edge, the top (001) and the side (111) facets of the probed nanocrystal, respectively.

\begin{figure*}[ht!]
  \centering
  \includegraphics[width=0.9\linewidth]{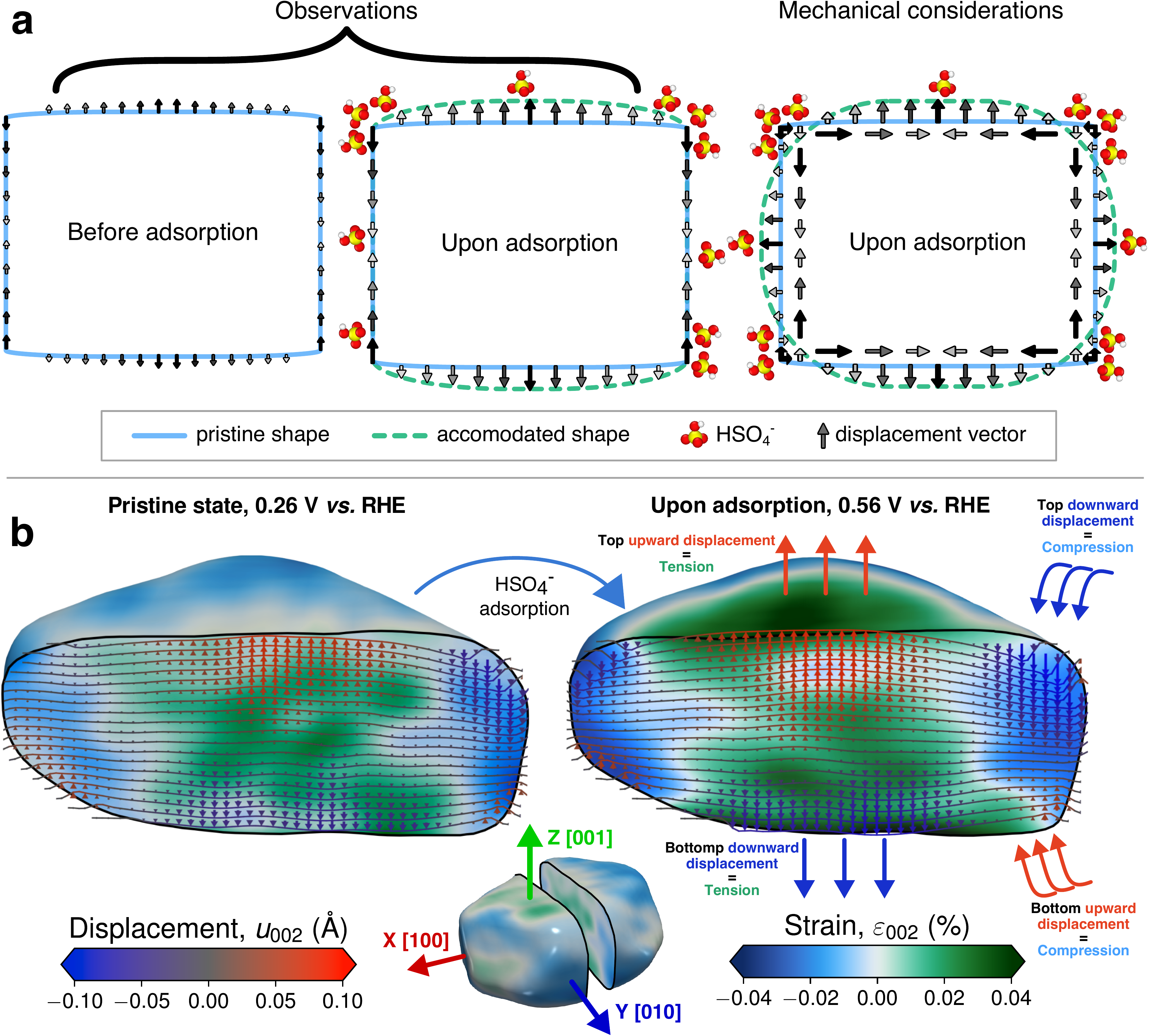}
  \caption{\textbf{Mechanical accommodation of a Pt nanoparticle upon bisulfate ion adsorption.} Panel \textbf{a} shows a 2D schematic of the response of a model shape due to the mechanical constraints induced by a Pt-HSO\textsubscript{4}\textsuperscript{-} adsorption. The left side corresponds to the observed behaviour allowed by the probing of the \textbf{002} Bragg peak. On the right side, we propose an interpreted view by considering a Poisson effect and therefore suggest that the in-plane compression and out-of-plane tension are the manifestation of the same adsorption phenomenon. On panel \textbf{b}, the curves and arrows represent the experimental displacement $\textit{u}_{002}$ of the bulk lattice parameter. The blue-to-red colour bar encodes the displacement. In the background, the blue-to-green colour bar corresponds to the derived strain $\varepsilon_{002}$. To avoid any misleading interpretation, the colour bars are perceptually uniform (see Colorcet collection \cite{noauthor_colorcet_nodate}). The length of the arrows exaggerates the displacement by a factor 22 for visibility. See Supplementary Figure 14 for other cross sections.}
  \label{fig:quiver}
\end{figure*}

Converting the experimentally calculated lattice displacement vectors into chemically useful local strain maps needs to consider the orientation of each facet on the particle surface. A downward displacement at the top of a crystal, towards the center of mass, is compressive, while the same downward displacement on the bottom of the crystal, away from the center of mass is tensile.  Because the BCDI dataset includes a single diffraction peak, it measures strain only along the [002] direction. This complicates the analysis, as chemically equivalent but directionally strained \{001\} facets exhibit different apparent displacement, depending on their relative orientation (see Supplementary Figures 10 and 11). This issue can fortunately be corrected through the well-known Poisson effect, where elastic distortion of a solid in one direction produces an opposing distortion perpendicular to the applied force (Figure \ref{fig:quiver}a). Out-of-plane tensile strain observed on the (001) and $(00\overline{1})$ facets therefore coincides with an in-plane compressive strain perpendicular to the [001] direction (inward displacement). Compressive in-plane strain on the (010), $(0\overline{1}0)$, (100) and $(\overline{1}00)$ facets implies these regions also experience a tensile out-of-plane strain. Of course, MSS/DFT modelling can account for strain in all directions simultaneously. Cross sections of the experimental projected displacement ($\textit{u}_{002}$) and strain ($\varepsilon_{002}$) fields at 0.26 V and 0.56 V are displayed in Figure \ref{fig:quiver}b. The coloured curves represent the local lattice displacement along the [001] direction as a deviation from the average crystal lattice\cite{robinson_coherent_2009}. The strain derived from the displacement is plotted for each pixel and superimposed to the displacement field using a blue-to-green colour bar. While bisulfate ions preferentially adsorb onto the corner and edge atoms to induce a local contraction, they also adsorb to \{001\} facets, creating in-plane compression and out-of-plane tension (Figure \ref{fig:quiver}b). Interestingly, the top and bottom facets experience similar strain response, while no adsorption takes place at the bottom facet since this is where the NP is bonded to the support. This can be rationalised by propagation of the strong edge and corner compressive strains propagating across the surface, which must be mechanically accommodated.


In summary, \textit{in situ} electrochemical BCDI was employed for the first time to probe the potential-dependent strain evolution of a single Pt nanoparticle. BCDI reveals adsorbate-induced surface strain initiating at edge and corner sites, and 3D, heterogeneous strain fields spreading across the faceted surface. A detailed understanding of facet-dependent mechanics, how strain fields create cross-talk between different active sites, and the influence of specific adsorption are critical for designing strain-engineered catalysts. Theoretical calculations not only explain the mechanistic origin of the strain fields, but demonstrate extensibility beyond this model system towards almost any practical industrial catalyst based on metallic nanoparticles, commonly found in energy conversion and storage systems.

\section*{Methods}
\paragraph*{Bragg coherent diffraction imaging} The BCDI measurements were performed with a photon energy of 13 keV ($\lambda =$ 0.95 \AA) at the ID01 beamline of the European Synchrotron Radiation facility (ESRF). The beam size was focused down to 690 nm (horizontally) $\times$ 630 nm (vertically) using Be compound refractive lenses. The electrochemical cell was mounted on a Physik Instrumente Mars \textit{xyz} piezoelectric stage with a lateral stroke of \SI{100}{\micro\metre} and a resolution of 2 nm, sitting on a hexapod that was mounted on a (3+2 circle) goniometer. Liquid electrolyte was continuously pumped through the cell during all measurements by a peristaltic pump. The diffracted beam was recorded with a 2D Maxipix\cite{ponchut_maxipix_2011} photon-counting detector (pixel size of \SI{55}{\micro\metre} $\times$ \SI{55}{\micro\metre}) positioned on the detector arm at a distance of 0.87 m. We measured the \textbf{002} Pt Bragg reflection in three dimensions by rotating the particle around the Bragg angle (16.12$^{\circ}$) through 1$^{\circ}$, in steps of 0.0125$^{\circ}$, with a counting time of 0.2 seconds/point and a total flux of about 1.4x10$^{12}$ counts/s/$\mu$m$^2$. The detector was positioned at an out-of-plane angle of $\sim$28$^{\circ}$. The time-scale of one BCDI measurement is \textit{ca.} 10 minutes.

\paragraph*{\textit{In situ} electrochemical measurements}
All glassware, electrodes, volumetric flasks, and components of the \textit{in situ} BCDI cell were first cleaned in 50 \% v/v solution of H\tsc{2}SO\tsc{4} (Merck, Suprapur 96 wt.\%) / H\tsc{2}O\tsc{2} (Carl Roth, 30 \% w/w) at least 12 hour, and then thoroughly washed with ultrapure water (Millipore, 18.2 $M\ohm$ cm, 1-3 ppb total organic compounds) before being used. The \textit{in situ} BCDI cell consisted of a polyetheretherketone (PEEK) body sealed by a \SI{6}{\micro\metre}  Mylar film. The electrochemical experiments were performed using a SP300 Biologic potentiostat and automatic Ohmic drop compensation. The working electrode was a 30 nm dewetted Pt film deposited onto a 6 mm glassy carbon cylinder (Sigradur grade G from Hochtemperatur-Werkstoffe GmbH). A Pt wire and a miniature Ag/AgCl/3.4 M Cl\tSc{-} were used as counter electrode and reference electrode, respectively. All electrode potentials are corrected relative to the reversible hydrogen electrode (RHE). Prior to the experiments, the electrolyte was deaerated with argon (Ar $>$ 99.999 \%, Messer). Then, 10 cyclic voltammograms between 0.023 and 1.24 V were performed at 50 mV s\tSc{-1} in Ar-purged 0.05 M H$_2$SO$_4$ solution.

\paragraph*{Phase retrieval algorithm} The reconstructed Bragg electron density and phase were obtained using the PyNX package\cite{favre-nicolin_fast_2011,favre-nicolin_pynx_2020}. Phase retrieval was carried out on the raw diffracted intensity data. The initial support, which is the constraint in direct space, was estimated from the auto-correlation of the diffraction intensity. A series of 1000 Relaxed Averaged Alternating Reflections (RAAR\cite{luke_relaxed_2004}), plus 400 Hybrid Input Output steps (HIO) as well as 300 Error-Reduction (ER\cite{gerchberg_practical_1972-1,fienup_reconstruction_1978}) steps, including shrink wrap algorithm\cite{marchesini_x-ray_2003}, were used. The phasing process included a partial coherence algorithm to account for the partially incoherent incoming wave front\cite{clark_high-resolution_2012}. To ensure the best reconstruction possible, we first selected the 15 best reconstructions (with lowest free Log-Likelihood\cite{favre-nicolin_free_2019}) from 40 with random phase starts. Then, 5 reconstructions were selected from the 15 by selecting the reconstructions with the lowest value of the standard deviation of the electron density. Finally, we performed the decomposition into modes from the last 5 reconstructions\cite{favre-nicolin_free_2019}. The reconstruction was then corrected for refraction and absorption using the bcdi package\cite{jerome_carnis_2022_5963911}. After removing the phase ramp and phase offset (Note that the phase is averaged over the whole particle. The mean value provides the phase offset.), the data was finally interpolated onto an orthogonal grid for ease of visualisation.

\paragraph*{Facet analysis} We have used the Facet Analyser plugin of the ParaView software, which is an automated 3D facet recognition algorithm that allows to detect the voxels at the surface of the particles and to index their facets by tuning parameters like sample size, angle uncertainty, splat radius or minimum relative facet size\cite{grothausmann_automated_2012}. The algorithm is based on the analysis of the probability distributions of the orientations of triangle normals of mesh representations of the particles. From the output list (voxels and associated global orientation) of the facet recognition algorithm, we have developed a Python script to determine the \textit{hkl} Miller indices of each facet as well as their average lattice displacement, strain and associated standard deviation.

\paragraph*{Definitions of strains.} The “surface strain” is the strain averaged over all surface voxels. Due to the size of the voxels, “surface strain” does not refer to distortion of the lattice parameter just within the outermost surface atoms but rather holds information averaged over the 20 outermost atomic layers. Similarly, “strain at edge and corner atoms” refers to the strain averaged over all voxels defining the edge and corner regions of the particle, “facet strain" to strain averaged over all surface voxels that are not edge/corner voxels, “bulk strain” to the strain averaged over all voxels that are fully surrounded by other voxels, and “global strain” to the strain averaged over all voxels of the NP.

\paragraph*{Molecular statics simulations}

In order to model accurately the experimental NPs, several materials simulation tools were used. We have used the Facet Analyser plugin of the ParaView software, which is an automated 3D facet recognition algorithm that allows to detect the voxels at the surface of the particles and to index their facets \cite{facetanalyser}. We then used the nanoSCULPT tool \cite{nanosculpt} to build realistic structures from the ParaView exported VTI files, for further atomistic simulations. Four nanocrystals were built with the following properties:

\begin{center}
\begin{tabular}{|c|c|c|c|c|}
    \hline
    Ratio & 1:2 & 1:5 & 1:10 & 1:20 \\ \hline
    Size (in nm) & 120$\times$115$\times$55 & 48$\times$46$\times$22 & 24$\times$23$\times$11 & 12$\times$11.5$\times$5.5 \\ \hline
    Number of atoms & 26 487 072  & 1 695 240 & 211 914 & 26 363 \\
    \hline
\end{tabular}
\end{center}

To obtain accurate and realistic relaxed configurations which reproduce as faithfully as possible the displacement fields measured in the experimental particle, MS simulations were carried out with the open-source Large-scale Atomic/Molecular Massively Parallel Simulator (LAMMPS) \cite{lammps, lammps2}. The interaction between atoms were modelled with different embedded-atom model (EAM) Pt potentials which predict slightly different elastic properties and surface energies, parameters that are essential to model accurately the experimental displacement and strain fields. We used the EAM potentials developed by Foiles \textit{et al.} \cite{foiles}. The NPs were relaxed at 0 Kelvin using a conjugate gradient algorithm to obtain the equilibrium displacement field. In order to allow a quantitative comparison with the experimental data, the three-dimensional diffraction patterns were calculated by summing the amplitudes scattered by each atom with its phase factor, following a kinematic approximation:
$$ I(\textbf{q}) = \left| \sum_j f_j (\textbf{q}) e^{-2 \pi i \textbf{q} \cdot \textbf{r}_{j}} \right| $$

where \textbf{q} is the scattering vector, $f_j$(\textbf{q}) and \textbf{r}\textsubscript{j} are respectively the atomic scattering factor and position of atom j. The computation was performed with a GPU using the PyNX \cite{pynx} scattering package, which considerably speed up the calculation of 3D diffraction patterns. The reciprocal volume over which the calculation was performed was selected based on the target voxel size in the real space data (same number of voxels in all simulations). Finally, the real space displacement and strain fields, u\textsubscript{002} and $\varepsilon$\textsubscript{002} were derived from the complex sample density $\rho$(r). The latter was obtained by performing a simple inverse Fourier transform of the scattered amplitude $\tilde{A}(\textbf{q})$:
$$ \tilde{\rho}(\textbf{r}) = \rho(\textbf{r})e^{2 \pi i \textbf{g} \cdot \textbf{u}(\textbf{r})} = FT^{-1}[\tilde{A}(\textbf{q})]$$

The projection of the displacement field $\textbf{g}_{002} \cdot \textbf{u}_{002}$ and the corresponding strain field $\varepsilon_{002} = \displaystyle\frac{\partial  {u}_{002}}{\partial  \textit{z}_{[001]}}$ could then be directly compared to the measured experimental strain field.

\paragraph*{Density functional theory simulations}

DFT calculations were performed using the Vienna Ab Initio Simulation package (VASP) \cite{vasp1, vasp2, vasp3}, within the projector-augmented wave method (PAW), for the description of the interaction between the valence electrons and the ionic core. The GGA-PBE functional was used, with DFT-D3 corrections \cite{dftd3} to account for possible van der Waals interactions. We consider atomic valences to be 3s\textsuperscript{2}3p\textsuperscript{4} (S), 2s\textsuperscript{2}2p\textsuperscript{4} (O) and 5d\textsuperscript{9}6s\textsuperscript{1} (Pt). Total energies were minimised until the energy differences were less than 1 $\times$ 10\textsuperscript{-6} eV between two electronic cycles. Atomic structures were relaxed until the Hellmann-Feynman forces were as low as 0.02 eV/\AA. Calculations were performed using a 450 eV cutoff energy and the following $\Gamma$-centered Monkhorst-Pack grids were used for the k-points meshes :

\begin{center}
\scalebox{0.85}{
\begin{tabular}{|c|c|c|c|c|c|c|}
    \hline
    System & Bulk Pt & Slab (001) & Slab (111) & Edge & Corner & NP \\ \hline
    k-points grid & 14$\times$14$\times$14 & 14$\times$14$\times$1 & 14$\times$14$\times$1 & 11$\times$3$\times$1 & 5$\times$5$\times$1 & 1$\times$1$\times$1\\ \hline
    Void thickness (\AA) & N.A. & 20 & 20 & 30 & 30 & 20 \\ \hline
    Size (in \AA) & 3.92 & 20 & 20 & 14 & 16 & 30$\times$30$\times$12\\ \hline
    Number of atoms & 4 & 24 & 40 & 92 & 242 & 573\\
    \hline
\end{tabular}
}
\end{center}

Those parameters were chosen to achieve a precision for the total energy lower than 0.1 meV/atom. In this work, the approach used to compute the adsorption of HSO\textsubscript{4} molecules onto Pt surfaces is the typical symmetric slab mode, wherein a supercell of the crystal oriented to expose its (hkl) surface is generated, and atoms are removed from a portion of the supercell to create a vacuum (thick symmetric slabs of 20 \AA, void thickness of 20 \AA). For each system, the two last atomic layers were fixed to bulk parameter to simulate a bulk structure. No ions (HSO\tsc{4} is considered, not HSO\tsc{4}\tSc{-}) nor implicit or explicit solvation were considered in this work. The adsorption energies were calculated as follow : $E_{ads} = E_{slab} - E_{surf} - E_{mol}$, where $E_{slab}$ is the energy of the slab (molecule on the surface), $E_{surf}$ is the energy of the surface without the adsorbed molecule and $E_{mol}$ is the energy of the lone molecule. A negative adsorption energy corresponds to a favorable adsorption configuration. The atomic strain tensor ($\overline{\overline{\varepsilon}}_{DFT}$) for each atom in the slab was extracted using the atomic strain modifier in OVITO \cite{ovito}, considering a cut-off radius of 3.09 Å. This value corresponds to the first minimum of the pair distribution function for Pt, \textit{i.e.}, halfway between the first and the second shell of neighbours. These strain fields could then be directly compared to the measured experimental strain field by calculating the simulated $\varepsilon_{002}$ component of the strain tensor :  $\varepsilon_{002} = \overline{\overline{\varepsilon}}_{DFT} \cdot \overline{n}_{002} \cdot \overline{n}_{002}$,  $\overline{n}_{002}$ being the unit vector pointing in the [002] (or [001]) direction. 



\noindent\textbf{Acknowledgements}\\
The authors are grateful to ESRF synchrotron for allocating beamtime, and to the ID01 beamline staff for the excellent support during the measurements. This project received funding from the European Research Council (ERC) under the European Union’s Horizon 2020 research and innovation programme (grant agreement No. 818823). The authors thank Pr. Philippe Sautet and Dr. Thierry Deutsch for the fruitful discussion on the DFT simulations.\\

\noindent\textbf{Author contributions}\\
M.-I. R. and F.M. conceived and designed the experiments. L.G. developed the electrochemical cell and performed the Pt/GC synthesis. C.A., I.M., M.D., N.L., S.L., T.S., F.M. and M.-I.R. performed the \textit{in situ} BCDI measurements.  C.A.,  C.C. and M.-I.R. analysed the data. C.C. performed the MSS and DFT calculations. C.A. wrote the first version of the manuscript. C.A., C.C., I.M., J.E., F.M. and M.-I. R. revised the manuscript.\\

\noindent\textbf{Competing interests}\\
The authors declare no competing interests. \\

\noindent\textbf{Data availability}\\
When published the data will be uploaded to the CXI database: https://www.cxidb.org. \\

\noindent\textbf{Supplementary information}\\
Supplementary information is available for this paper.\\

\noindent\textbf{Correspondence and requests for materials} should be addressed to C.A. (clement.atlan@esrf.fr), F.M. (frederic.maillard@lepmi.grenoble-inp.fr) or M.-I.R. (mrichard@esrf.fr).


\end{document}